\documentclass[aps,twocolumn,prl,showpacs]{revtex4}
\usepackage{graphicx,amssymb,amsmath}
\usepackage{graphicx}
\usepackage{dcolumn}
\usepackage{bm}

\begin{document}

\title{A Hot Electron Thermocouple and the Diffusion Thermopower of Two-Dimensional Electrons in GaAs}

\author{W.~E. Chickering$^1$, J.~P. Eisenstein$^1$, and J.~L. Reno$^2$}

\affiliation{$^1$Condensed Matter Physics, California Institute of Technology, Pasadena, CA 91125
\\
$^2$Sandia National Laboratories, Albuquerque, NM 87185}

\date{\today}

\begin{abstract}
A simple hot electron thermocouple is realized in a two-dimensional electron system (2DES) and used to measure the diffusion thermopower of the 2DES at zero magnetic field.  This hot electron technique, which requires no micron-scale patterning of the 2DES, is much less sensitive than conventional methods to phonon-drag effects.  Our thermopower results are in good agreement with the Mott formula for diffusion thermopower for temperatures up to $T \sim 2$ K.

\end{abstract}

\maketitle

The thermoelectric properties of low-dimensional electronic systems provide information about carrier transport that is complementary to that obtained from ordinary charge transport.  For example, in an ordinary Drude metal the electrical conductivity $\sigma$ is simply proportional to the momentum scattering time $\tau$. In contrast, the diffusion thermopower $S_d$ depends upon both $\tau$ and its energy derivative $d\tau/dE$ \cite{AM}. Additional motivation for measuring thermopower comes from its close connection to the entropy per particle in the low dimensional system.  While this connection has been long appreciated for non-interacting electrons \cite{Obraztsov}, it has also been found to hold in strongly interacting, disorder-free cases at high magnetic field, notably the half-filled lowest Landau level \cite{Nigel}.  Very recently it has been suggested that thermopower may even reflect the excess entropy associated with non-abelian quasiparticle exchange statistics \cite{Kun}.  Finally, beyond these very fundamental motivations there is also the simple fact that thermopower, harnessed in a humble thermocouple device, provides a very effective way to measure temperature.  

In semiconductor-based two-dimensional electron systems experimental access to the diffusion thermopower and the important information it contains has been limited by the parasitic effects of phonons \cite{review,Fletcher}.  In a typical experiment the needed temperature gradient is established by applying heat to one end of a bar-shaped sample while the other end is thermally ``grounded''. The overwhelming majority of the applied heat is transported by phonons.  The resulting phonon wind exerts a drag force on the electron gas which leads to a thermoelectric voltage independent of that arising from the diffusion thermopower of the electrons themselves.  This phonon-drag thermopower, $S_{ph}$, can exceed $S_d$ by more than an order of magnitude.  Only by going to very low temperatures ($T\lesssim 0.2$ K) can the diffusion component of the thermopower be observed in such experiments \cite{Ying}.

\begin{figure}
\includegraphics[width=3.3in, bb=0 0 234 312]{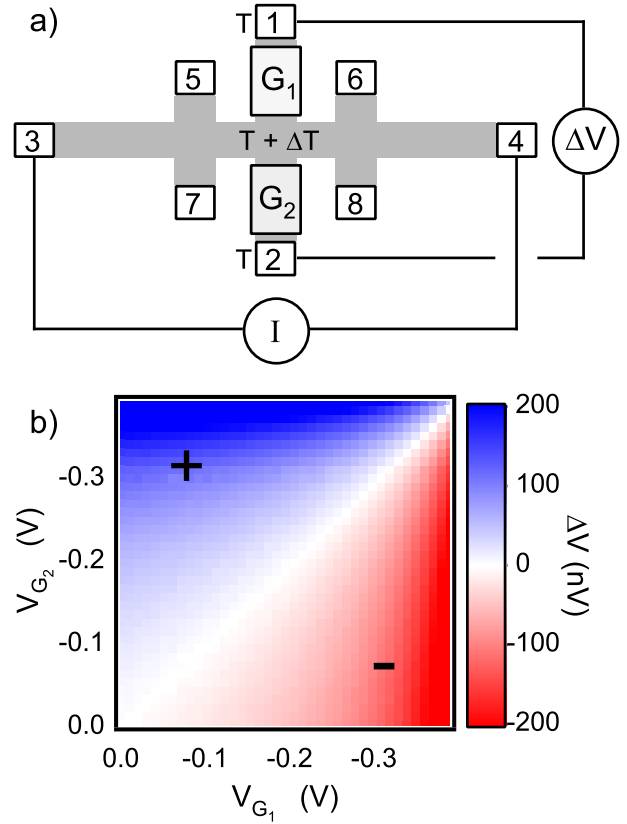}
\caption{\label{}(color online) a) Layout of simple 2D hot electron thermocouple device. Numbered squares are ohmic contacts, light gray rectangles are surface gates G$_1$ and G$_2$. b) Color intensity plot of thermoelectric voltage arising from Joule heating the 2D electron gas.}
\end{figure}
In this paper we report new measurements of the thermopower $S$ of a two-dimensional electron system (2DES) in a GaAs/AlGaAs heterostructure.  A simple hot electron technique is employed which greatly reduces the contribution of phonon drag to the measured thermopower \cite{hotelectron}.  As a result we are able to show that the magnitude and the temperature and density dependences of $S$ are in good agreement with the Mott formula \cite{Mott} for the diffusion thermopower $S_d$ for temperatures up to about $T \sim 2$ K, an order of magnitude higher than in previous measurements \cite{Ying}.  For $T \gtrsim 3$ K we observe increasing deviations from the Mott formula, most likely due to phonons.

The 2DES employed here has a nominal density and mobility of $n=1.6\times 10^{11}$ cm$^{-2}$ and $\mu = 3.3 \times 10^6$ cm$^2$/Vs, respectively, at low temperatures. The 2DES is patterned into the device geometry illustrated in Fig. 1(a). A 1 mm-long, 60 $\mu$m-wide bar has ohmic contacts at each end for driving current along it. Three arms extend away from each side of the central bar and terminate at ohmic contacts. The two opposing arms in the center of the device are overlaid by gate electrodes G$_1$ and G$_2$ on the sample surface; these two arms function as a 2DES thermocouple.  The remaining arms and ohmic contacts are used as voltage probes enabling measurements of the resistance $R$ of the 2DES in the central region of the device. 

The ohmic contacts, which consist of a NiAuGe alloy diffused into the sample, enable electrical measurements and also serve to thermally anchor the electron gas in their immediate vicinity to the lattice temperature $T_l$.  However, away from the contacts the electron gas can be easily heated out of equilibrium with the lattice. For example, driving current $I_{3,4}$ between contacts 3 and 4 will raise the temperature $T_e$ of the 2DES in the center of the device above $T_l$.  Since no magnetic field is applied, no voltage difference $\Delta V = V_1-V_2$ will appear between the opposing contacts 1 and 2. Not only will there be no voltage due to the longitudinal or Hall resistivities of the 2DES, the symmetry of the device will also prevent a thermoelectric voltage difference from developing in spite of the elevated electron temperature in the center of the device.  In order for there to be a thermoelectric voltage between contacts 1 and 2, the 2DES in their respective mesa arms must have different thermopowers.  This is easily achieved by electrostatically modifying the 2DES densities in the arms, $n_1$ and $n_2$, using gate electrodes G$_1$ and G$_2$. In this way arms 1 and 2 constitute a 2D hot electron thermocouple.

Figure 1(b) shows, in a color intensity plot, the thermoelectric voltage $\Delta V$ observed in response to $I_{3,4}=2~\mu$A at $T_l = 1$ K for a range of voltages $V_{G1}$ and $V_{G2}$ applied to the two gates.  The figure clearly demonstrates that $\Delta V \approx 0$ along the diagonal where $V_{G1} = V_{G2}$ and $n_1 = n_2$. Away from this equal density condition $\Delta V$ is non-zero.  Note that $\Delta V$ changes sign as the equal density diagonal is crossed. This is expected since, as in all thermocouples, it is the {\it difference} in the thermopowers of the two arms that counts.

Throughout this paper the drive current $I$ used for heating is ac (at typically $f=13$ Hz).  However, the voltage difference $\Delta V$ is lock-in detected at $2f$, i.e. at {\it twice} the drive frequency.  This is appropriate because Joule heating scales with the square of the current \cite{currentdep}. This technique has the advantage of being insensitive to ordinary resistive voltage drops, as might result from small asymmetries in the device, which would appear at the fundamental drive frequency, $f$.

While the data in Fig. 1(b) are qualitatively consistent with the hot electron thermocouple model we have proposed, they are not immediately useful for a quantitative determination of the thermopower.  Beyond the conversion from gate voltage to density in the thermocouple arms (which is readily accomplished via simple magneto-conductance measurements), there is the more difficult issue of knowing the electron temperature $T_e$ in the heated region.  For this purpose precision measurements (using a modified Zair-Greenfield bridge \cite{ZG}) of the resistance $R$ of the bar's central region are made using contacts 5 and 6 (or 7 and 8).  For temperatures above about $T \approx 0.8$ K the temperature dependence of the resistance is sufficient to allow a calibration of the temperature rise of the 2DES due to Joule heating induced by the drive current, $I$.

In brief, the measurement scheme used in this experiment is as follows.  The GaAs/AlGaAs heterostructure sample and the ohmic contacts to the 2DES within it are in excellent thermal contact with the cold-finger of a cryostat \cite{DR} whose temperature $T_l$ is accurately measured and regulated using a calibrated resistance thermometer \cite{Cernox} and a commercial temperature controller \cite{LR700}. The resistance $R$ of the 2DES in the central region of the mesa is first measured with a drive current $I$ small enough to ensure that negligible heating of the 2DES occurs.  The lattice temperature $T_l$ is then decremented by an amount $\Delta << T_l$, causing $R$ to change slightly. This change in $R$, which reflects the cooling of the 2DES, is then eliminated by increasing $I$ and thereby heating the electron gas out of equilibrium with the crystal lattice and the ohmic contacts.  At this point a known \cite{subtle} temperature difference $\Delta T$, within the electron gas, has been established between the central region of the device and the ohmic contacts at the ends of the thermocouple arms.  After biasing the gates to produce known electron densities, $n_1$ and $n_2$, in the two thermocouple arms, the thermoelectric voltage difference $\Delta V$ (measured at $2f$) between the ohmic contacts at the ends of those arms is recorded \cite{error}. Additional measurements of $\Delta V$, with the lattice temperature decremented by $2 \Delta$, $3 \Delta$, $etc.$ are used to improve accuracy and ensure that $\Delta V$ remains linear in $\Delta T$ (in all cases the maximum $\Delta T$ is less than $0.1~T_l$).  The slope of $\Delta V$ $vs.$ $\Delta T$ represents $\Delta S \equiv S_1-S_2$, the difference in the thermopowers of the two arms.

According to the Mott formula \cite{Mott}, the diffusion thermopower $S$ of a 2DES that behaves as a simple Drude metal is

\begin{equation}\label{eq.1}
S=-\frac{\pi^2k_B}{3e}\frac{T}{T_F}(1+\alpha)
\end{equation}
where $T_F$ is the Fermi temperature (proportional to the density, $n$) and $\alpha$ reflects the energy and thus density dependence of the electronic momentum scattering time $\tau$: $\alpha \equiv(n/\tau)\frac{d\tau}{dn}$.  At low temperatures, where $\tau$ is dominated by impurity scattering, $\tau$ and hence the parameter $\alpha$ are very nearly temperature independent.  In typical GaAs-based 2DESs $\alpha$ ranges from about $\alpha \sim0.7$ to 1.5, depending on the details of the impurity potential in the sample \cite{dassarma}. Density-dependent resistivity measurements on the 2DES in an adjacent chip of the same MBE wafer used for the present thermopower studies reveal that at low temperatures $\alpha \approx 0.92 \pm 0.05$ in our device over the density range of interest, $1.57>n>0.32 \times 10^{11}$ cm$^{-2}$.  Hence, Eq. \eqref{eq.1} predicts that at low temperatures $S$ is proportional to $T$ and inversely proportional to density, $n$ \cite{2DMIT}.

\begin{figure}
\includegraphics[width=3.2in, bb=0 0 254 216]{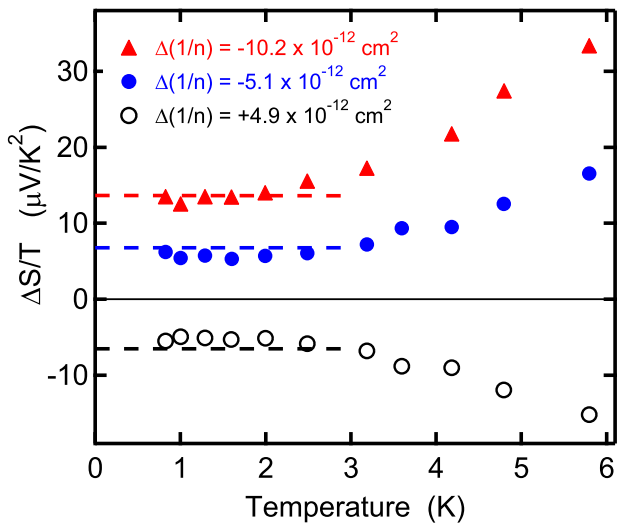}
\caption{\label{}(color online) $\Delta S/T$ $vs.$ $T$ for $\Delta(1/n) = -10.2$, -5.1, and $+4.9\times 10^{-12}$ cm$^2$.  Dashed lines are predictions of Mott formula, Eq. \eqref{eq.1}.}
\end{figure}
Figure 2 shows representative determinations of $\Delta S/T$ $vs.$ $T$, for three different electron density combinations in the two thermocouple arms.  Below about $T = 2$ K, $\Delta S/T$ is independent of $T$ while at higher temperatures it begins to rise.  The temperature independence of $\Delta S/T$ for $T\lesssim2$ K is consistent with the Mott formula for the diffusion thermopower but contrasts sharply with the much stronger temperature dependences observed previously at these relatively high temperatures.  This difference reflects the much reduced importance of phonon drag in the present measurements.

The three data sets shown in Fig. 2 are qualitatively consistent with the expected density dependence of the thermopower; $\Delta S/T \propto \Delta(1/n) \equiv n^{-1}_1-n^{-1}_2$. The sign of $\Delta S/T$ changes with the sign of $\Delta (1/n)$ and the magnitude of $\Delta S/T$ clearly increases with $\Delta (1/n)$.  To examine this dependence more carefully, Fig. 3 displays the dependence of $\Delta S$ on $\Delta (1/n)$ at $T = 1$ and 2 K.  Although neither data set is perfectly linear in $\Delta (1/n)$, the deviations are relatively small.  

Having established that our measured thermopower has the temperature and density dependence expected of thermal diffusion in the 2DES, we turn to the magnitude of the effect.  The dashed horizontal lines in Fig. 2 and the solid diagonal lines in Fig. 3 are the predictions of Eq. \eqref{eq.1} for the various data sets shown. Since we have separately measured the parameter $\alpha$, Eq. \eqref{eq.1} contains no adjustable parameters.  While the overall agreement between theory and experiment is clearly quite good, there are small systematic deviations. For example, Fig. 2 shows that the magnitude of the measured  thermopower at $\Delta (1/n) = \pm 5 \times 10^{-12}$ cm$^2$ falls below the Mott result by about 20 percent for $T \lesssim 2$ K.  Similar deviations are apparent in the density dependences shown in Fig. 3.  Although the origins of these deviations are so far unknown, one systematic effect that may be important deserves mention. The electron temperature determinations used here are based on resistivity measurements using voltage probes that straddle the thermocouple arms.  As such, these measurements offer an average of $T_e$ within the central region of the device.  The electron temperature in this region is determined both by the ability of the 2DES to lose energy to phonons but also by the conduction of heat through the 2DES to the several ohmic contacts.  Simulations of these heat transfer processes in our device suggest that the average $T_e$ inferred from the resistivity measurements {\it exceeds} that sensed by the thermocouple junction by typically 10 percent.  While correcting for this effect would reduce the discrepancy between our data and the Mott formula, we defer doing so until a more thorough analysis of the effect is done.

\begin{figure}
\includegraphics[width=3.2in, bb=0 0 252 216]{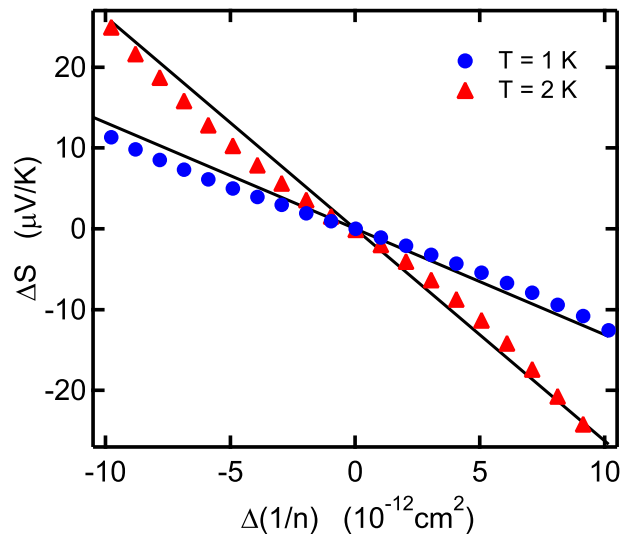}
\caption{\label{}(color online) $\Delta S$ $vs.$ $\Delta(1/n)$ at $T = 1$ and 2 K.  Solid lines are predictions of Mott formula, Eq. \eqref{eq.1}.}
\end{figure}
As Fig. 2 shows, the magnitude of the observed $\Delta S/T$ begins to exceed the Mott prediction as the temperature is increased beyond about $T \sim 3$ K.  We speculate that this is due to phonon drag effects.  At these high temperatures phonon emission by the heated 2DES is significant. The phonons emitted from the central region of the device will exert a drag force on the electrons in the thermocouple arms and thereby enhance the measured thermoelectric voltage.  We stress that while this drag process is qualitatively similar to that observed in prior 2DES thermopower measurements, it is far weaker in the present instance.  Since our technique relies on direct heating of the 2DES, relatively little heat is needed to produce a given electron temperature gradient, $\nabla T_e$.  In contrast, previous experiments, which rely on heating of the crystal lattice, require much larger heat inputs to create the same $\nabla T_e$.  This distinction is directly evident in the factor-of-ten increase in the temperature range ($T \lesssim 2$ K $vs.$ $T \lesssim 0.2$ K) over which the measured thermopower is dominated by electron diffusion instead of phonon drag.

As a final observation, we note that the present thermopower measurements are limited to $T \gtrsim 0.8$ K only because calibration of the electron temperature via the resistivity becomes extremely difficult at lower temperatures.  However, it is certainly possible to measure thermoelectric voltages due to electron heating at lower temperatures.  If one assumes the Mott relation remains valid, then it should not be difficult to resolve a 3 mK temperature rise at $T_e = 30$ mK \cite{estimate}.  We emphasize that the present 2DES thermocouple directly measures the \emph{electron} temperature.  Low dimensional electron systems are notoriously difficult to cool into the few mK temperature range and disequilibrium between the lattice and electron gas is commonplace.  Conventional thermometers, which are usually much more strongly coupled to the lattice phonons than to the electron system, are of limited use in detecting such a disequilibrium.  We therefore believe that the hot electron thermocouple technique presented here will find application to a variety of thermal measurements on low dimensional electron systems at low temperatures.

In conclusion, we have devised a simple hot electron thermocouple and used it to measure the diffusion thermopower of 2D electrons in a GaAs/AlGaAs heterostructure.  We find good agreement with the Mott formula for the thermopower for temperatures below about $T = 2$ K.  This technique promises to provide electron thermometry down to very low temperatures.  

We thank Gil Refael, Sankar Das Sarma, and Ady Stern for helpful discussions.  This work was supported via DOE grant DE-FG03-99ER45766 and Microsoft Project Q.


\begin{references}
\bibitem{currentdep} As expected, for small heating currents the magnitude of the observed thermoelectric voltage is proportional to $I^2$.  At higher currents the temperature rise in the heated region becomes sub-linear in the power dissipated owing to the strong temperature dependence $\dot{Q} \propto (T_e^5-T_l^5)$ of the electron cooling rate due to acoustic phonon emission.
\bibitem{DR} A dilution refrigerator circulating only a small fraction of its normal $^3$He-$^4$He mixture provides excellent thermal stability at the relatively high temperatures ($0.6 <T_l<6$ K) of interest here.
\bibitem{error} Altering the density of the 2DES in the thermocouple arms can have a small effect on the degree of electron heating due to the current flowing in the central bar of the device. Our estimates suggest that this effect can be safely ignored.
\bibitem{estimate} With $\Delta S/T \approx 15 \mu$V/K$^2$ (as shown in Fig. 3 for $\Delta (1/n) =-10 \times 10^{-12}$ cm$^2$), $\Delta T_e =3$ mK at $T_e=30$ mK yields a thermoelectric voltage $\Delta V \approx 1$ nV.  This is readily detectable.
\end{references}
\end{document}